# INTEGRATED 3D SOUND INTENSITY SENSOR WITH FOUR-WIRE PARTICLE VELOCITY SENSORS

D.R. Yntema, J.W. van Honschoten, R.J. Wiegerink

Transducers Science and Technology group, MESA+ Institute for Nanotechnology, University of Twente

P.O.Box 217, 7500 AE Enschede, the Netherlands

*Abstract-* A new symmetrical four-wire sensor configuration has resulted in a fully integrated sound intensity sensor with significant lower noise floor and smaller size than its predecessors. An integrated sound pressure sensor was further miniaturized by using a folded "back chamber" at both sides of the chip.

I. INTRODUCTION

Acoustic sound fields consist of a combination of pressure change and particle velocity. A three dimensional particle velocity sensor together with a pressure sensor provides all information regarding the sound field at the measurement spot. The particle velocity sensor consists of two small platinum sensor (2 μm wide, 300 nm in height and 1.5 mm long) wires placed at a distance of 350 μm from each other. When these wires are heated by an electrical current and a flow component is present in the plane of the wires and perpendicular to the length axis of the wires this will result in a temperature difference between the wires. This difference in temperature will on its turn result in a difference in wire resistance and therefore in an output signal, which is proportional to the particle velocity level, see reference [1]. In this paper a similar sensor but with four wires instead of two is presented. When more sensors are used in a different direction a more-dimensional particle velocity sensor is made.

A three dimensional particle velocity sensor has been developed earlier; see [2] and [3]. Differences between previous designs and the sensor configuration described here are a smaller size and a lower noise. Also a pressure transducer is included, and compared to earlier versions it has a much smaller size and good predictable response. The noise performance of this sensor is good enough to be used in commercial applications when packaged further.

Furthermore a (signal processing) method for electronic adjustment of the directivity of the sensor is presented.

II. A 3D PARTICLE VELOCITY SENSOR

A 3D particle velocity sensor with two times four wires placed in parallel and at neighboring distances of 250 μm (see figure 1) is developed.

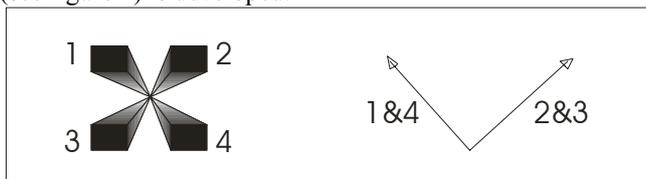

Fig. 1. The sensor wires as seen in the length axis with sensitivity direction.

Two diagonally placed wires are used as one sensor with wire distance of 350 μm, so each four-wire configuration results in two sensor output signals corresponding to orthogonal particle velocity directions. For each sensor there is now an additional heat source in the form of the other pair of sensor wires, and this result in a higher sensitivity. Therefore, both a smaller size and better performing sensor is realized. To make a three dimensional type two of these sensors are places in a silicon chip as shown in the schematic figure 2 below.

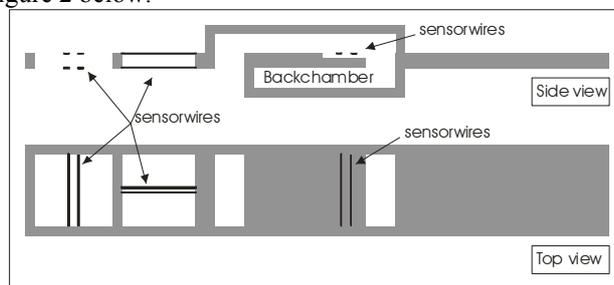

Fig. 2. The design schematically represented.

By linearly combining the two sensor signals the sensitive direction can be adjusted. This can be clarified with the help of figure 3.

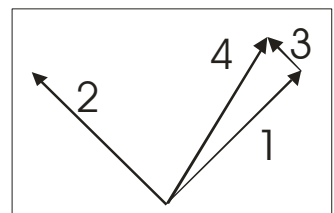

*Figure 3, directivity correction method*

When a part of the (analogue) signal '2' is added to signal '1', then the direction of sensitivity shifts by '3' to direction '4'. This technique is possible in the acoustic frequency range for two reasons; the sensors are almost identical and are virtually at the same place. With the technique it is possible to rotate the direction of sensitivity over 360 degrees, but here it is used to correct a deviation between the expected (desired) direction of sensitivity and the experimentally measured direction.

III INTEGRATED PRESSURE MICROPHONE

The integrated pressure sensor does not contain moving parts like a diaphragm, but instead an acoustic pressure to particle velocity converter [3, 4] is realized. By using this converter sound pressure can be measured by a particle velocity sensor, allowing the use of the same fabrication process as with the particle velocity sensors. The particle velocity sensor is positioned inside an acoustic 'back chamber' which is folded on the top and bottom side of the sensor. In figure 2 this is represented schematically. Compared with the sensor described in [3] the pressure sensor is smaller in width, but somewhat longer. Since the sensor has to be mounted anyway and mounting is normally done in the length-axis, larger sensor length is less unfavorable than sensor width.

The response of the sensor is a function of the acoustic impedance of the back chamber and the response of the particle velocity sensor as described in [3] and [4].

## IV FABRICATION

Fabrication is done on both sides of a 250 μm thick double side polished silicon wafer. The wafer thickness defines the distance between the sensor wires. The wafer is covered with an LPCVD (Low Pressure Chemical Vapor Deposition) low stress silicon nitride layer of 200 nm thickness. Next a layer of Chromium/Platinum is applied on top of a patterned photo resist layer; with a lift off bath the photo resist (with the metal to be removed on top) is removed and the sensor wires are left. Next the holes in the silicon nitride layer are etched by RIE (Reactive Ion Etching), at the places where the wires should be free hanging. In KOH (Potassium Hydroxide) etchant the silicon nitride layer acts as a mask layer, but at places where the layer is etched away the silicon is etched by the KOH. For the pressure sensor two sensor wires are used, and additionally two caps are etched out of another wafer. The caps are mounted on top of the sensor chip using epoxy glue. Since the caps are of equal outer size and the glue has a low surface tension the caps are self-aligning.

Figure 4 shows a complete mounted 3D sensor with integrated pressure sensor, the sensor wires of the particle velocity sensor are well visible. One of the pressure caps of the pressure sensor is just below them.

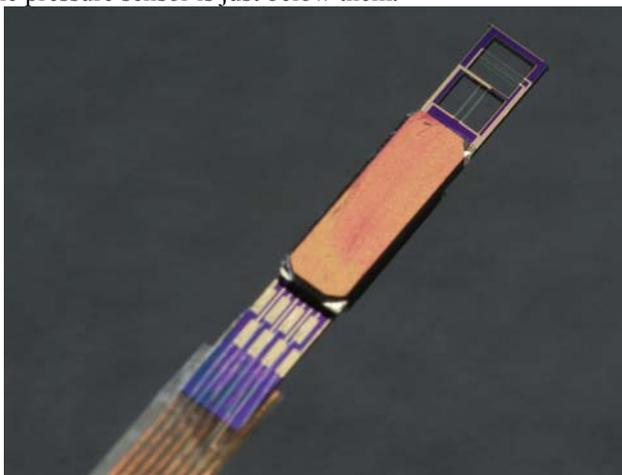

Fig. 4. Photo of the mounted device with mounted pressure back chamber.

## V MEASUREMENT RESULTS

To verify that this sensor is directionally sensitive for particle velocity a polar plot is measured. For the 3D particle velocity sensor the polar pattern is measured and plotted in Figure 5, 'no correction'. Clearly the sensor is directional, but an (almost) frequency independent deviation of 15 degrees between desired and measured sensitivity direction exist. This can be compensated for with different placement of sensor wires, but also electronically by combining the sensor signals as mentioned above. With an electronic correction circuit the polar plot can be adjusted to match exactly 45 degrees as shown in figure 4 'corrected'.

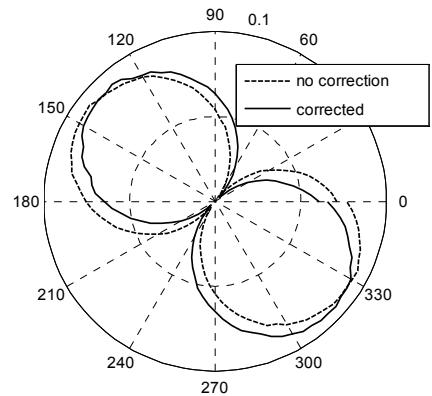

Figure 5; polar pattern of particle velocity sensors, relative sensitivity measured with steps of 3.75 degree. Rotation is along the length-axis. The required directions are 45 and 135 degrees; deviation is a constant 15 degrees. Sensitivity direction can be corrected. (See 'corrected'). The polar plots are measured at a frequency of 600 Hz, but for frequencies measured up to 5 kHz the polar plot was consistent.

To investigate how much better the performance of a four wire sensor is compared with a regular two wire design the sensitivity and noise level are measured with two wires and with four wires. Dividing the electrical noise level in [Volts] by its sensitivity [V/(m/s)] gives the apparent acoustical selfnoise in [m/s]. Selfnoise levels of the four wire sensor are compared with the two wire type in figure 6. Clearly the four wire sensor has a better selfnoise performance.

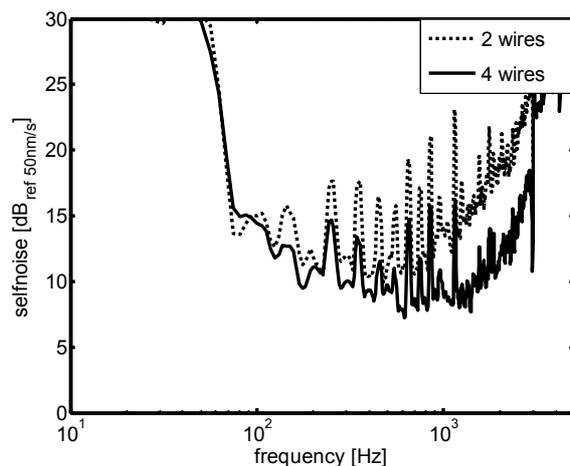

Figure 6; Selfnoise levels with 2 wires and 4 wires powered. Especially at higher frequencies the four wire sensor has a lower noise level.

The pressure microphone has to be pressure sensitive only. To confirm this, the sensor is placed in a standing wave tube.

At certain frequencies and positions in the tube there will be a maximum in pressure level. At these places and frequencies the particle velocity is minimal and vice versa. For proper functioning as a pressure sensor, the sensor is required to give maximal signal at pressure maxima and minimal signal at particle velocity maxima (pressure minima). In figure 7 the measurement together with a particle velocity sensor is shown. Indeed at particle velocity maxima the signal from the pressure device is minimally and at particle velocity minima the signal is maximally. Concluding from this the sensor is only pressure sensitive. Furthermore, since pressure is a scalar the response of the sensor when rotated must be omni directional. Measurements in figure 8 confirm this, there is no directivity effect.

the standard fabrication process for the particle velocity sensors a pressure sensitive sensor has been made.

ACKNOWLEDGMENT

The authors wish to thank the Dutch Foundation for Technology STW for financial support.

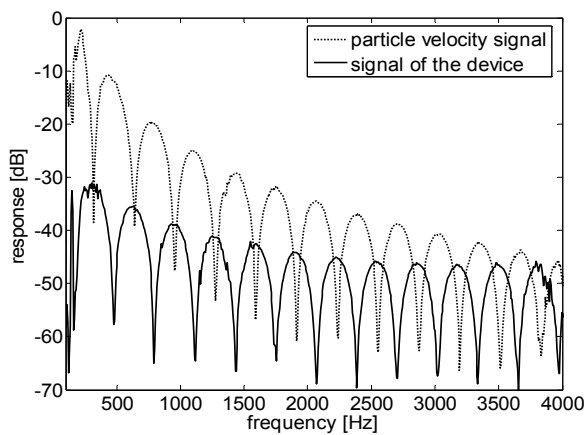

*Figure 7; Pressure sensitivity proof of the device; at pressure maxima there is a particle velocity minimum, and vice versa. The device is only pressure sensitive*

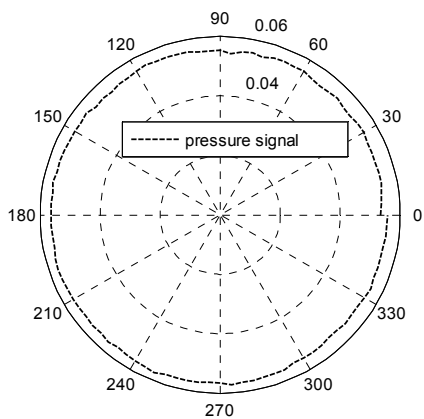

*Figure 8; Pressure sensor polar plot measured at 700 Hz, but results are constant over frequency 50 to 4000Hz. The sensor is omni directional.*

## VI CONCLUSION

A three-dimensional sound intensity probe consisting of four particle velocity sensors and one pressure sensor in a silicon substrate was developed. The sensor size is smaller than its predecessors, and the four wire sensor design has better performance in terms of selfnoise level. Together with